\documentclass{aipproc}

\usepackage{graphicx} 
\usepackage{amssymb}
\usepackage{amsmath}

\layoutstyle{6x9}
\def\eps{\epsilon}
\def\al{\alpha}

\begin{document}

\title{Coherent $\nu-N$ scattering and the search for physics
beyond the standard model}
\author{J. Barranco}{
address={Instituto de    F\'{\i}sica, Universidad Nacional Aut\'onoma
  de M\'exico,
    Apdo. Postal 20-364,  01000 M{\'e}xico D.F., \ M{\'e}xico.
}
}



\begin{abstract}
We focus in future proposals to measure coherent neutrino-nuclei scattering 
and we show that such kind of experiments are very sensitive to nonstandard 
neutrino interactions with quarks. First in a model independent parametrization
 and then we focused in particular models such as leptoquarks and models with 
extra neutral gauge bosons and with R-parity breaking interactions. We show 
that in all these three different types of new physics it is possible to
obtain competitive bounds to those of future collider experiments. For the 
particular case of leptoquarks we found that the expected sensitivity to the 
coupling and mass for most of the future experimental setups is quite better 
than the current constraints. 
\end{abstract}


\maketitle


\section{The coherent neutrino-nucleus cross section}
There are few predictions made by the Standard Model (SM) that have not yet been observed, most of the times due to experimental difficulties. One of these non-observed predictions is the so called coherent neutrino-nucleus scattering \cite{Freedman:1973yd}. In this process a neutrino scatters coherently not only the nucleons but the nucleus itself. The coherent scattering requires momentum transfer, $q$, small compared with the inverse nucleus size, $q \le 1/R$, $R$ the nuclear radius. This condition is well satisfied for almost every typical atom, in the case of neutrino energies coming from reactors, artificial sources, the sun or supernovas. Neglecting radiative corrections the cross section is
\begin{eqnarray}
\frac{d\sigma}{dT}&=&\frac{G_F^2 M}{2\pi}\left\{
(G_V+G_A)^2+\left(G_V-G_A\right)^2\left(1-\frac{T}{E_\nu}\right)^2-
\left(G_V^2-G_A^2\right)\frac{MT}{E_\nu^2}
\right\}\,,\label{diff:cross:sect}
\end{eqnarray}
where $M$ is the mass of the nucleus, $T$ is the recoil nucleus energy, $E_\nu$ is the incident neutrino energy and the axial and vector couplings are
\begin{eqnarray}
\label{GV}
G_V&=& 
\left[g_V^p Z+g_V^nN\right]
F_{nucl}^V(q^2)\,,\\
G_A&=& 
\left[g_A^p \left(Z_+-Z_-\right)+g_A^n\left(N_+-N_-\right)\right]
F_{nucl}^A(q^2)\,.
\label{GA}
\end{eqnarray}
$Z$ and $N$ represent the number of protons and neutrons in the
nucleus, while $Z_{\pm}$ ($N_\pm$) stands for the number of protons
(neutrons) with spin up and spin down respectively.
The vector and axial nuclear form factors, $F_{nucl}^V(q^2)$ and
$F_{nucl}^A(q^2)$, are usually assumed to be equal and of order of
unity in the limit of small energies, $q^2\ll M^2$.
The SM neutral current vector couplings of neutrinos with
protons, $g_V^p$, and with neutrons, $g_V^n$, are 
\begin{eqnarray}
&&g_V^p=\rho_{\nu N}^{NC}\left(
\frac12-2\hat\kappa_{\nu N}\hat s_Z^2
\right)+
2\lambda^{uL}+2\lambda^{uR}+\lambda^{dL}+\lambda^{dR},\nonumber\\
&&g_V^n=-\frac12\rho_{\nu N}^{NC}+
\lambda^{uL}+\lambda^{uR}+2\lambda^{dL}+2\lambda^{dR}\,.
\label{vcouplings}
\end{eqnarray}
Here $\hat s_Z^2=\sin^2\theta_W=0.23120$, $\rho_{\nu N}^{NC}=1.0086$,
$\hat\kappa_{\nu N}=0.9978$, $\lambda^{uL}=-0.0031$,
$\lambda^{dL}=-0.0025$ and $\lambda^{dR}=2\lambda^{uR}=7.5\times10^{-5}$
are the radiative corrections given by the 
PDG~\cite{Yao:2006px}. The axial contribution can be neglected as can be seen from eq. (\ref{diff:cross:sect}) since the ratio of axial to vector contribution is expected to be of the order $1/A$, $A$ the atomic number. The spin-zero cross section of electron 
neutrino scattering off nucleus in the low energy limit, $T \ll E_{\nu}$ is 
\begin{equation}
\frac{d\sigma}{dT}(E_\nu,T)=\frac{G_F^2 M}{\pi}
\left(1-\frac{M T}{2E_\nu^2}\right)\left[
Z (g_V^p)+N (g_V^{n})\right]^2.\label{CS}
\end{equation}
\begin{center}
\begin{figure}
\includegraphics[angle=270,width=0.5\textwidth]{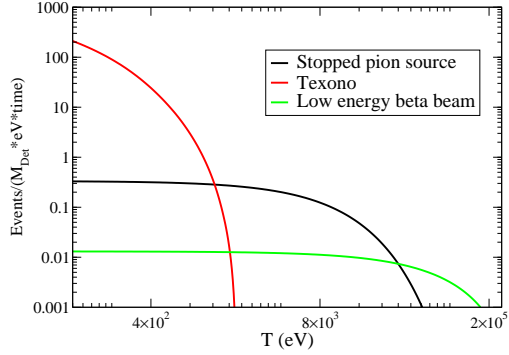}
\caption{Events vs. energy for three different experimental proposals}
\end{figure}
\end{center}
The number of expected events, neglecting for a moment the detector efficiency 
and resolution, can be estimated as
\begin{equation}
N_{\rm{events}}=t\phi_0\frac{M_{\rm{detector}}}{M}
\int\limits_{E_{min}}^{E_{max}}dE_\nu
\int\limits_{T_{th}}^{T_{max}(E_\nu)}dT
\lambda(E_\nu)\frac{d\sigma}{dT}(E_\nu,T)\,,
\label{Nevents}
\end{equation}
with $t$ the data taking time period, $\phi_0$ the total neutrino
flux, $M_{\rm{detector}}$ the total mass of the detector,
$\lambda(E_\nu)$ the normalized neutrino spectrum, $E_{max}$ the
maximum neutrino energy and $T_{th}$ the detector energy threshold. One
can see that the maximum nuclei's recoil energy 
depends on nucleus mass $M$
through the relation $T_{th}^{max}=2E_\nu^{max^2}/(M+2E_\nu^{max})$. 
This is the biggest experimental challenge to defeat in order to detect this 
process. For example, reactor 
neutrino with $E_{\nu}^{max}=10$MeV produces a maximum nucleus recoil 
for Germanium of $12$KeV. 
Nevertheless, experiments that look for direct dark matter searches 
have improved
the detectors at very low threshold. 
Different proposals have been suggested to detect the coherent $\nu-N$ 
scattering that differs each other depending on the neutrino source and the
detector properties.  

\begin{table}[!t]  
        \begin{tabular}{cccc}
            \hline     
 Experiment  & $M_0$ &Expected events/yr& systematic error estimate \\
\hline 
 Texono, $E_{th}=$400 eV    & 1 kg, Ge    & 3790  & 2 \%  \\ 
 Texono, $E_{th}=$100 eV    & 1 kg, Ge    & 25196 & 2 \%  \\ 
 Beta beam, $E_{th}=$15 keV & 1 ton, Xe    & 1390  & 2 \%  \\ 
 Beta beam, $E_{th}=$5 keV  & 1 ton, Xe    & 5309  & 2 \%  \\ 
 Stopped pion, $E_{th}=$10 keV  & 100kg, Ne    & 627   & 5 \%  \\  
 \hline
        \end{tabular}  
    \caption{Expected events for different experimental setups}
    \label{detectors}
\end{table}

{\bf The experimental proposals:}\\
{\bf TEXONO:} Texono collaboration has recently started a 
research program towards
the measurement of neutrino-nuclei coherent scattering by using
reactor neutrinos and 1~Kg of an ``ultra-low-energy'' germanium detector
(ULEGe)~\cite{Wong:2005vg} with a threshold as low as 100~eV and we have
consider also a threshold of  400~eV. 
We can consider that the error will be dominated by the reactor power, 
its fuel composition, and the
anti-neutrino spectrum. We assume that these uncertainties will give
an approximate error of 2\%~\cite{Huber:2004xh} and one year of data taking.\\ 
{\bf Stopped pion source:} Another proposal for detecting the coherent neutrino-nucleus
scattering considers the use of another source of neutrinos, a 
stopped pion source (SPS),
such as the Spallation Neutron Source at Oak Ridge National
Laboratory ~\cite{Scholberg:2005qs}.
The neutrino spectra are well known.  Here we will consider only the total
delayed flux ($\nu_e$~$+$~$\bar{\nu_\mu}$) that comes from pion decay. 
We assume a total flux of
$\sim10^7\nu$~s$^{-1}$~cm$^{-2}$. Among different possible detector materials
such as Ar, Ge or Xe, we will concentrate on the noble gas detector,
$^{20}$Ne, of typical mass about $100$~kg with a data taking time scale from
one to several years and a threshold of $10$~keV.\\
{\bf Low energy Beta-Beams:} The usage of accelerated radioactive nuclei to produce a well known
flux of neutrinos -- beta beam -- was proposed
in~\cite{Zucchelli:2002sa}. It was shown soon afterwards that low
energy beta beams open new possibilities to study neutrino
properties~\cite{Volpe:2003fi} and, recently, a neutrino-nuclei coherent
scattering experiment using neutrinos from low energy beta beams was
discussed~\cite{Bueno:2006yq}.  In particular we base our analysis on 
the beta-beam experiment discussed
in~\cite{Serreau:2004kx,Bueno:2006yq}.
The number of expected neutrinos observed in this proposal is
\begin{equation}
\label{dNevdt}
  N_{events}^{\beta-beam}=t\, g\,\tau\, n\,h\times
  \int_0^\infty dE_\nu\,\Phi_{tot}(E_\nu)\,\sigma(E_\nu)\,,
\end{equation}
where $t=1$~year is the data taking time, $n$ is the number of target nuclei
per unit volume, $\sigma(E_\nu)$ is the relevant neutrino-nucleus
cross-section. We considered the case of a ton of Xe as a
target and a factor $\gamma=14$ for $^6$He ions as described in
Ref.~\cite{Bueno:2006yq}.  As for the threshold energy, we consider both the
realistic threshold of $15$~keV where background events are negligible as well
as the very optimistic $5$~keV threshold that, according to the same
reference, will give a bigger number of events if background could be
subtracted, though at present there is no technology capable of dealing with
such a background.  The total neutrino flux through detector is given in
\cite{Serreau:2004kx}.

\section{The non standard interaction}
We can go a step further and look for new physics in the neutrino sector. 
New Physics such as non standard interactions (NSI) that can be included adding to the 
Standard Model (SM) Lagrangian the effective Lagrangian \cite{Davidson:2003ha} 
\begin{equation}\label{lagrangiannsi}
-{\cal L}^{eff}_{\rm NSI} =
\eps_{\alpha \beta}^{fP}{2\sqrt2 G_F} (\bar{\nu}_\alpha \gamma_\rho L 
\nu_\beta)
( \bar {f} \gamma^\rho P f ) 
\end{equation}
where $f$ is a first generation SM fermion: $e,u$ or $d$, and $P=L$ or $R$.
That kind of NSI Lagrangian appear in a natural way for example when it is 
added mass to the neutrinos  
\cite{Schechter:1980gr}
and also appear from super-symmetric theories \cite{Barger:1989rk}. 
So, all new physics will be included
in the effective parameter $\eps_{\al \beta}^{fP}$ and the goal is to 
constrain such parameters using the neutrino data.
But the solar and atmospheric data allow big values on such NSI parameters and 
even worst, the appearance of new ``dark'' solutions \cite{Miranda:2004nb}. It turns now relevant to really understand correctly the NSI parameters if a good
determination in the mixing parameters is wanted. Experiments at low energy 
offer an excellent opportunity for the NSI study, because the no dependence in
neutrino mixing parameters. In particular, the coherent neutrino-nucleus
scattering can set strong bounds in NSI parameters because the enhancement 
in the number of detectable events as can be seen in Table \ref{detectors}.

{\bf The NSI in $\nu-N$ scattering:} 
The coherent neutrino-nucleus scattering cross section after the 
addition of Lagrangian \ref{lagrangiannsi} is easily obtained with the 
change in the coupling constant 
$g_v^p \to g_v^p+2\varepsilon_V^u+\varepsilon_V^d$ and 
$g_v^n \to g_v^n+\varepsilon_V^u+2\varepsilon_V^d$ in eq. (\ref{CS}). 
The expected number of events depends now in the NSI parameters 
$N_{Events}(\varepsilon_V^d, \varepsilon_V^u)$ and it is possible to 
constraint the NSI parameters by doing a $\chi^2$ analysis taking the SM
as central value. 
Taking only one parameter at a time and just as illustrative
we show the possible constraint obtained with a Texono-like experiment with
two different threshold energies in Table \ref{texonolike}. The sensitivity for this 
kind of experiments is very competitive with future $\nu-$factories.

\begin{table}
\begin{tabular}{cccc}
\hline
Present Limits & $^{76}$Ge $T_{Th}=400$eV&$^{76}$Ge $T_{Th}=100$eV& $\nu$Factory\\
\hline
$-0.5 < \eps_{ee}^{dV} < 1.2  $ & $|\eps_{ee}^{dV}|<0.003$ &$|\eps_{ee}^{dV}|<0.001$&$|\eps_{ee}^{dV}|<0.001$\\
\hline
$-1.0 < \eps_{ee}^{uV} < 0.61  $ & $|\eps_{ee}^{uV}|<0.002$ &$|\eps_{ee}^{uV}|<0.001$&$|\eps_{ee}^{uV}|<0.002$\\
\hline
\end{tabular}
\caption{Constrains in NSI parameters with Texono-like experiment at 90\% C.L. Present limits and $\nu~$Factory sensitivity are taken from \cite{Davidson:2003ha}. For details see
\cite{Barranco:2005yy}} \label{texonolike}
\end{table}
Once we constrained the NSI parameters, it is natural to ask if the
bounds in the effective parameter can be translated into constrains in
specific models. Here we did in three different realizations of theories 
beyond the standard model:Heavy neutral vector bosons $Z'$, Leptoquarks and
supersymmetric theories with $R$-parity breaking terms.\\  

\section{NSI in  different extensions of SM}

{\bf Heavy neutral vector bosons:} Extra $Z'$ bosons are predicted in
string inspired extensions of the SM, in left-right symmetric models,
in models with dynamical symmetry breaking, in "little Higgs" models
and in certain classes of theories with extra dimensions. In many of
these models it is expected that $Z'$ mass can be around TeV scale.
The present experimental lower limits to the neutral gauge boson mass
come from the Tevatron and LEP experiments~\cite{Yao:2006px}.
Forthcoming measurements at LHC will provide sensitivity to the $Z'$
mass up to 5~TeV~\cite{Dittmar:2003ir,Rizzo:2006nw}. The effect of this
$Z'$ boson will generate a neutral current that modifies the SM coupling
constants with a NSI parameters as expressed in Table \ref{couplings},where 
$c_\beta=\mbox{cos}\beta$, $s_\beta=\mbox{sin}\beta$ and $\gamma=(M_Z/M_{Z'})^2$
are the relevant parameters.  \\
{\bf Leptoquarks:} A leptoquark is a scalar or vector boson that couples to a lepton and
a quark. There are no such interactions in the SM, but they are
expected to exist in various extensions of the SM~\cite{Yao:2006px}, 
such as the Pati-Salam model~\cite{Pati:1974yy},
grand unification theories based on
$SU(5)$~\cite{Georgi:1974sy,Dorsner:2005ii} and
$SO(10)$~\cite{Fritzsch:1974nn} gauge groups and extended technicolor
models~\cite{Farhi:1980xs}. The leptoquark contribution effectively 
(in 4-fermion approximation) can be written as expressed in Table \ref{couplings} 
~\cite{Davidson:1993qk} where $\lambda_u$, $\lambda_d$ are couplings, $m_{lq}$ is leptoquark
mass. This parametrization is given for vector leptoquarks.  In the
case of scalar leptoquarks, our results should be multiplied by a
factor 1/2~\cite{Davidson:1993qk}.\\
\begin{table}
\begin{tabular}{cccc}
\hline
& Heavy neutral gauge boson $Z'$ & Leptoquarks & R-parity breaking terms \\
\hline
$\varepsilon^{uV}$ &$0$&$ \frac{\lambda_u^2}{m_{lq}^2}\frac{\sqrt{2}}{4G_F}$&0\\
\hline
$\varepsilon^{dV}$ &$-4\gamma\sin^2\theta_W \rho_{\nu N}^{NC}
\left({3 c_\beta \over 2 \sqrt{24}}+
{s_\beta \over 6}\sqrt{5 \over 8} \right){c_\beta \over \sqrt{6}} $& 
$\frac{\lambda_d^2}{m_{lq}^2}\frac{\sqrt{2}}{4G_F}$&$ \left({M_W^2 \over g^2}\right) 
\left({|\lambda^\prime_{1j1}|^2 \over m_{\tilde d_{jL}}^2}-{|\lambda^\prime_{11k}|^2 
\over m_{\tilde d_{kR}}^2}\right)$\\
\hline
\end{tabular}
\caption{Effective 4-fermion NSI parameter expressed in terms of the
relevant parameters for  three different models beyond the SM.}\label{couplings}
\end{table}
{\bf R parity breaking terms:} In supersymmetric theories, gauge invariance does not imply baryon
number (B) and lepton number (L) conservation and, in general, the so
called R-parity (defined as $R=(-1)^{3B+L+2S}$ where $S$ is the spin)
is violated. However, one has to keep the consistency with the non-observation of
fast proton decay. One may consider, for instance, the R-parity violating MSSM 
(imposing baryon number conservation) with 
a superpotential that contains the $L$- violating
terms~\cite{Barger:1989rk} $\lambda_{ijk}L_{L}^i L_{L}^j \bar E_{R}^k $
,$\lambda_{ijk}^\prime L_{L}^i Q_{L}^j \bar D_{R}^k $,
where we use the standard notation, $L_L, Q_L, \bar E_R$, and $\bar
D_R$ to denote the chiral superfields containing the left-handed
lepton and quark doublets and the right-handed charged-lepton and
$d$-quark singlets respectively; $i,j,k$ are generation indices. 
At low energies, the heavy Supersymmetry particles can be integrated out
and the net effect of the $R$-breaking interactions is to generate
effective 4-fermion operators involving the lepton and quark fields.
By considering the case where a single Yukawa coupling (with one
flavor structure) is much larger than the others, the effective
four-fermion operator generated by $L_L^i Q_L^j \bar D_R^k$ takes the
same form as in Eq. (\ref{lagrangiannsi}) with the couplings
~\cite{Barger:1989rk,Chemtob:2004xr} as expressed in Table \ref{couplings}.
\section{Results}
Our results are summarized in Table \ref{results}. For definiteness we have fixed some
parameters. The estimated sensitivity has been calculated for the detector's mass 
given in the proposal and the time=1 year of data taking. In case of $Z'$ we have estimated 
the sensitivity of each experimental proposal for the mass for the extra gauge boson in the
particular case where $cos\beta=1$, that is called the $\chi$-model. 
For  leptoquarks we have fixed the coupling $\lambda_q$ to the electroweak scale ($\lambda_q^2=4 \pi/137$) and the estimation are on the leptoquark mass at 90\% C.L. in GeV. Finally, for SUSY with $R$-parity breaking terms, the estimation is done over the difference in the coupling constants $|\lambda^\prime_{1j1}|^2/ m_{\tilde d_{jL}}^2-|\lambda^\prime_{11k}|^2 / m_{\tilde d_{kR}}^2$ with s-quark masses normalized to 100~GeV. We can see that in most of the cases we have competitive or as in the leptoquark case much better
sensitivity to the relevant parameters than the current constraints. For a more detailed discussion see \cite{Barranco:2007tz}.\\
{\bf Acknowledges} This work was partially supported by DGAPA-UNAM.
\begin{table}
\begin{tabular}{ccccccc}
\hline
&Texono$_{100 eV}$ &Texono$_{400 eV}$ & SPS & $\beta$-Beam$_{15 KeV}$ &$\beta$- Beam$_{5KeV}$& Current\\
\hline
$Z'$&792&722&479&619&725&680 (GeV)\\
Leptoquark& 894&805& 546& 684&805&298 (GeV)\\
$R$-Parity&0.0020&0.0025&0.005&0.003&0.0024&0.012\\
\hline
\end{tabular}
\caption{Constraint in the relevant parameters for each model}\label{results}
\end{table}



\end{document}